\documentclass[10pt, conference]{IEEEtran}
\IEEEoverridecommandlockouts

\usepackage{cite}
\usepackage{amsmath,amssymb,amsfonts}
\usepackage{algorithmic}
\usepackage{graphicx}
\usepackage{textcomp}
\usepackage{xcolor}
\usepackage{pgf-pie}
\usepackage{tikzscale}
\usepackage{pgfplots}
\def\BibTeX{{\rm B\kern-.05em{\sc i\kern-.025em b}\kern-.08em
    T\kern-.1667em\lower.7ex\hbox{E}\kern-.125emX}}
\begin{document}

\title{Accelerated Deep Lossless Image Coding with Unified Paralleleized GPU Coding Architecture\\
}

\author{
\IEEEauthorblockN{Benjamin Lukas Cajus Barzen}
\IEEEauthorblockA{\textit{NUE Research Group} \\
\textit{TU Berlin}\\
Berlin, Germany \\
b.barzen@tu-berlin.de}
\and
\IEEEauthorblockN{Fedor Glazov}
\IEEEauthorblockA{\textit{KGB Research Group} \\
\textit{TU Berlin}\\
Berlin, Germany \\
glazov@campus.tu-berlin.de}
\and
\IEEEauthorblockN{Jonas Geistert}
\IEEEauthorblockA{\textit{NUE Research Group} \\
\textit{TU Berlin}\\
Berlin, Germany \\
geistert@nue.tu-berlin.de}
\and
\IEEEauthorblockN{Thomas Sikora}
\IEEEauthorblockA{\textit{NUE Research Group} \\
\textit{TU Berlin}\\
Berlin, Germany \\
sikora@nue.tu-berlin.de}
}

\maketitle

\begin{abstract}
 
We propose Deep Lossless Image Coding (DLIC), a full resolution learned lossless image compression algorithm. Our algorithm is based on a neural network combined with an entropy encoder. The neural network performs a density estimation on each pixel of the source image. The density estimation is then used to code the target pixel, beating FLIF in terms of compression rate. Similar approaches have been attempted. However, long run times make them unfeasible for real world applications. We introduce a parallelized GPU based implementation, allowing for encoding and decoding of grayscale, 8-bit images in less than one second. Because DLIC uses a neural network to estimate the probabilities used for the entropy coder, DLIC can be trained on domain specific image data. We demonstrate this capability by adapting and training DLIC with Magnet Resonance Imaging (MRI) images. 

\end{abstract}


\section{Introduction} \label{intro}
When encoding data losslessly, one well studied and universally applicable concept is entropy coding. Considering data comprised of symbols, entropy coding can compress this data given a probability distribution over all possible symbols. Utilizing entropy coding, recent research of neural network based lossless image compression approaches the problem as one of density estimation \cite{pixel-cnn-1, pixel-cnn-2, pixel-cnn-3, practical-full-resolution-compression}. A neural network is trained to estimate a probability density function (PDF) for a target pixel. The target pixel is then encoded with entropy coding using the estimated PDF. More specifically, the network learns a PDF \begin{equation}
    P(x_i) = P(x_i|x_j, x_{j+1}, .., x_{k})
\end{equation} where $x_i$ is the target pixel and $x_{j}$ to $x_{k}$ is a collection of neighboring pixels.  Prominent implementations include PixelSnail\cite{pixel-cnn-1}, PixelCNN/PixelCNN++ \cite{pixel-cnn-2,pixel-cnn-3} and Practical Full Resolution Lossless Image Compression (L3C)\cite{practical-full-resolution-compression}. PixelCNN++, designed for image generation, trains a neural network to produce a parametrization for a mixture of logistics, based on preceding pixels in the image. The estimated mixture of logistics is then used to sample the next pixel in the image, but it can also be used to compress the pixel in an existing image \cite{pixel-cnn-3}. PixelCNN++ achieves 2.92 bits on average per RGB pixel (bpp)\cite{pixel-cnn-3}, but is limited in its practical use for image compression because of 18 to 80 minutes of coding time per 320x320 crop \cite[Table 4]{practical-full-resolution-compression}. 
L3C addresses this issue by first recursively extracting and downscaling feature maps from the target image, using a neural network. The pixels in the image are then predicted based on the topmost feature map using a second neural network. To make the topmost feature map available during decoding, it itself is recursively predicted and compressed using the lower-resolution feature maps. The smallest feature map is stored alongside encoded data as-is. Like PixelCNN++, the L3C network predicts a parametrization of a mixture of logistics and is based on a residual neural network architecture. L3C is able to use such a complex architecture and simultaneously drive down decoding time to approximately 200-370 milliseconds \cite[Table 4]{practical-full-resolution-compression}, because all mixture model parametrizations for all pixels are predicted at once. L3C's compression performance appears to be slightly inferior to FLIF's, while reducing the encoding time by a factor of seven but increasing the decoding time by a factor of three \cite[Table 1 and 4]{practical-full-resolution-compression}.  

We propose Deep Lossless Image Coding (DLIC), an alternative approach to parallelizing neural network based lossless image coding. We demonstrate that a fully connected neural network can achieve good prediction performance, while also being small enough for massive parallelization. Using this approach, DLIC outperforms all common image formats including HEVC Intra as well as decision-tree based format FLIF. 

\section{Description of the Coder} \label{description}
\subsection{Coder Design} \label{algorithm}

\begin{figure}[htbp]
\centerline{\includegraphics[width=21mm]{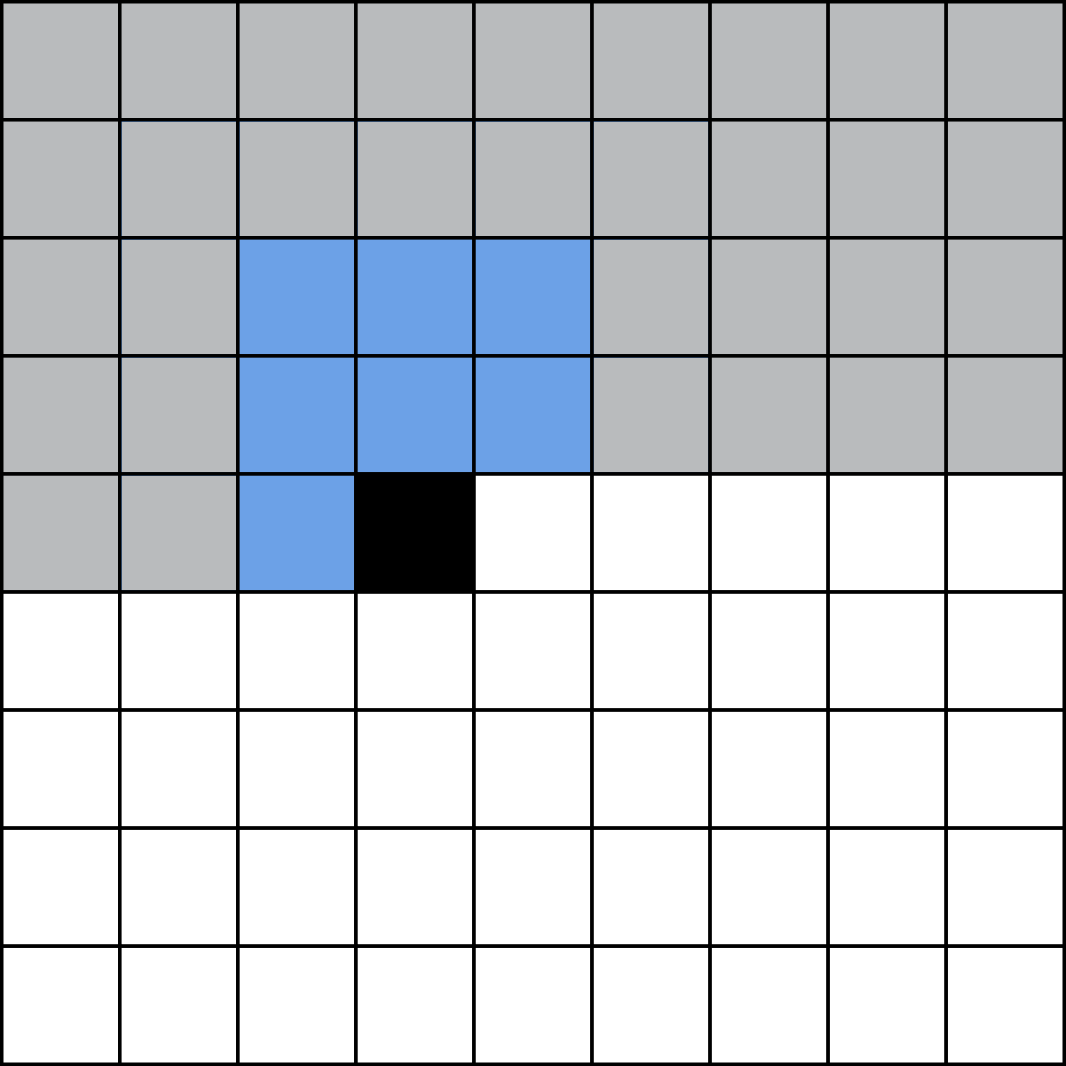}\hspace{10mm}\includegraphics[width=29mm]{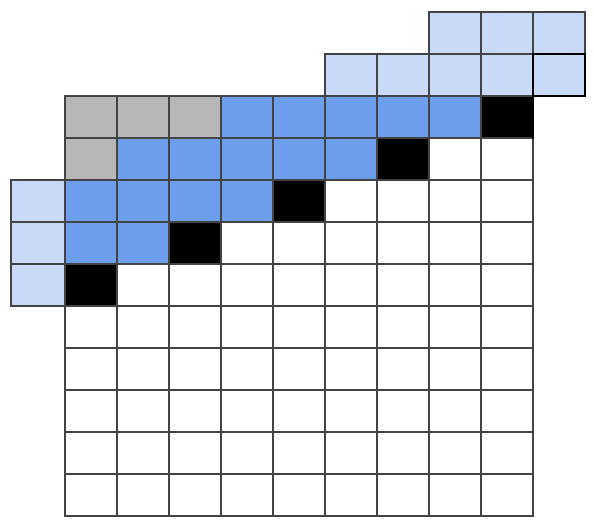}}
\caption{\textbf{Left:} Schematic example of a step in the decoding process of DLIC. The black pixel is the one we wish to decode, gray pixels are already decoded pixels, blue pixels are the priors for the black pixel.
\textbf{Right:} DLIC wavefront visualization. Blue Pixels are in the coding window of the black pixels and already de/encoded. The light blue pixels are not part of the original image and are substituted with dummy values.}
\label{image_overview}
\end{figure}

\begin{figure}[htbp]
\centerline{\includegraphics[width=24mm]{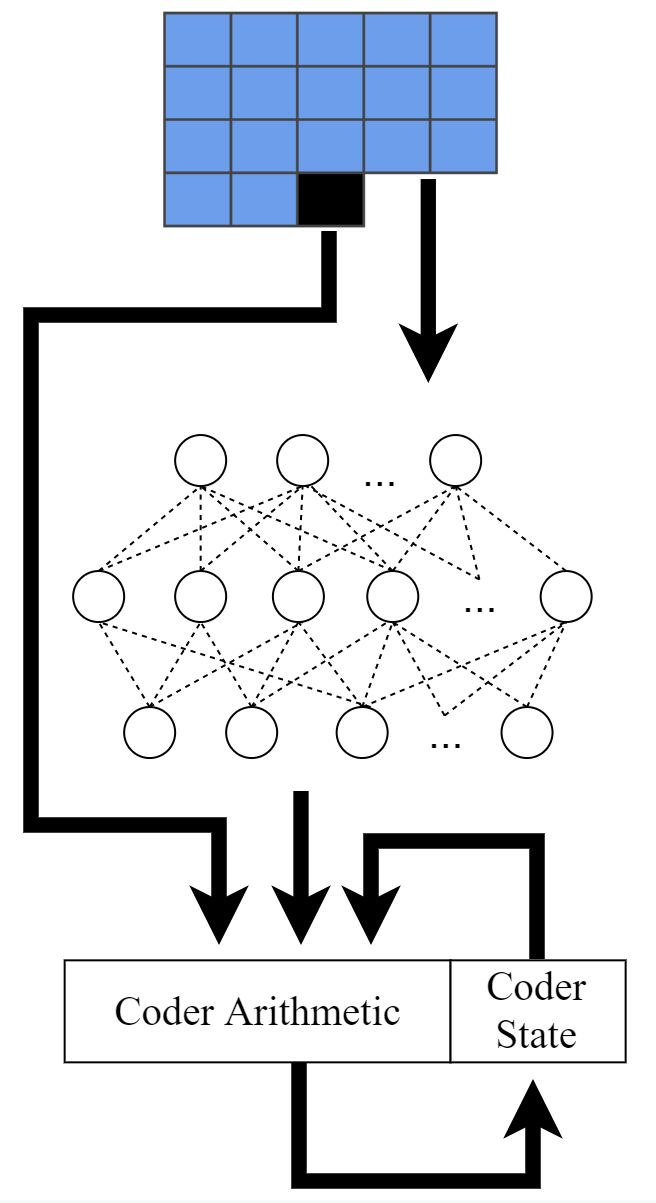}\hspace{10mm}\includegraphics[width=24mm]{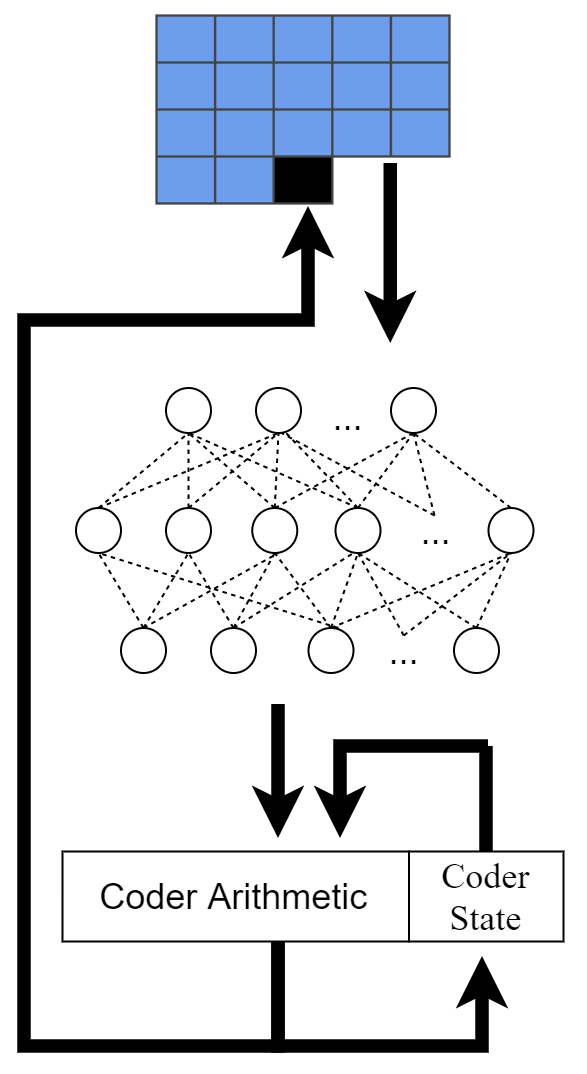}
}
\caption{\textbf{Left:} Schematic overview of our encoder consisting of (top to bottom) preprocessor which extracts pixel neighborhoods, a neural network with window size input variables and number of pixel values outputs, as well as a coder which encodes the pixel to be encoded (black) with the output of the neural network.
\textbf{Right:} Schematic overview of our decoder. Note that the black pixel is to be decoded and not part of the input.}
\label{decode_step}
\end{figure}

Deep Lossless Image Coding (DLIC) consists of a neural network performing density estimation, as well as an entropy coder to compress the pixel values. In each coding substep, the network performs a probability density estimation of a target pixel. The neural network is trained to perform the density estimation based on a neighborhood window consisting of pixels in the vicinity of the target pixel. To ensure that the decoder can reproduce the probability density estimation this neighborhood window is masked to only contain the already decoded pixels. Figure \ref{image_overview} (left) depicts an example of such a substep, with the neighborhood window colored blue and the target pixel colored black. The dimensions of the neighborhood window shown are not fixed, but any window shape can be trained and used. Figure \ref{decode_step} depicts the information flow when encoding (left) and decoding (right) a pixel. When decoding, the pixel is not know yet, but encoded in the coder state. To extract it, the neighborhood window is passed to the neural network, which generates a PDF over all possible target pixel values. The PDF is then passed to the entropy coder, which decodes the pixel from the coder state. The pixel can then be part of the next neighborhood window. When encoding, DLIC passes the neighborhood window of the target pixel to the neural network. This step is identical with the decoding process, and it is assured that the neighborhood window consist of the same pixels. The target pixel is then passed to the entropy coder alongside the PDF, and is subsequently encoded in the coder state.

\subsection{Implementation} \label{implementation}

We compartmentalized the whole coder system into three components:
\begin{itemize}
\item A preprocessor, which extracts neighborhoods out of the image (during encoding) and reassembles the image (during decoding).
\item A neural network which predicts this PDF, using the neighborhood as the input. 
\item An entropy coder, which en- and decodes symbols based on a PDF.
\end{itemize}

\subsubsection{Preprocessor}
While it is possible to achieve superior compression rates using neural network based models for lossless image compression \cite{pixel-cnn-1, pixel-cnn-2, pixel-cnn-3, practical-full-resolution-compression}, these coders often ignore real world time and hardware constraints. While \cite{practical-full-resolution-compression} suggests speeding up lossless compression with a more complex neural network and coder architecture, we developed a conceptually simple implementation that performs quickly by taking advantage of modern graphics cards and highly parallelizing the workload.

We use a principle similar to wavefront parallel processing (WPP) as defined in the HEVC standard \cite{hevc}. This allows us to decode and encode multiple pixels at the same time. Instead of traversing the image rowwise left to right during coding, we instead determine in each coding step which pixels we may code next. These pixels are then coded in parallel. The preprocessor, neural network and entropy coder all run in parallel on the graphics card for these pixels. The encoding wavefront is visualized in Figure \ref{image_overview} (right).

As a result of the parallelization, the neural network has to predict the probabilities for multiple pixels at the same time. Instead of a vector input consisting of one pixel neighborhood, it now expects a matrix input, with each row consisting of one pixel neighborhood. Note that in order to guarantee reproducible results, it is paramount that the same matrices enter the neural network in the encoding and decoding steps. This ensures that all rounding errors, inherent in floating point operations used in neural networks, are the same during encoding and decoding. This does not mean that we always need to en- and decode on the same graphics card. As long as the  precision of the floating point arithmetic is the same, the results are equivalent. 

The implementation of the preprocessor as well as the  implementation of the other two components, the entropy coder and the neural network, are also realized on the GPU. This has the advantage that we only need to communicate twice between the CPU and the GPU: Once to load the image onto the graphics card, and another time to retrieve the output of the coder. With this approach we keep the overhead of communicating between the CPU and the GPU to a minimum. 

\subsubsection{Neural Network Architecture}

Our proposed model is a dense network consisting of six dense layers with two optional pooling layers between the first and second and the third and fourth layer. The output layer consists of 256 neurons, each representing the probability of a 8-bit grayscale value. The output layer uses the softmax function as activation function. The number of neurons in the hidden layers varies between 128 neurons to 4096 neurons per layer, depending on targeted pixel depth and desired compression rate.  

\subsubsection{rANS: Range Asymetric Number System Coder}
\label{sec:rANS}

Asymmetric Number Systems were first suggested by Jarek Duda \cite{ransDuda} and are now the basis of a variety of coding schemes, most notably JPEG XL, Facebook ZSTD and Apple LZFSE\cite{ANSreview}. ANS based methods find widespread use because they can achieve a speedup by a factor of 20 over Arithmetic Coding while sustaining the compression ratio \cite{dudaarXiv}. We used the rANS variant for our coder.

As pointed out previously, the rANS Coder was also implemented on the GPU to eliminate inter CPU/GPU communication during coding. Since rANS works in a sequential fashion, but we wish to code multiple pixel values in parallel due to the wavefront approach, the rANS implementation consists of multiple coder instances. These coder instances code into/from multiple bitstreams, depending on the image dimensions and at most one per pixel row. In our CPU implementation, 64-bit arithmetic is used. In the GPU implementation, we opted for 32-bit arithmetic to avoid slowing down the coder on graphic cards that do not support native 64-bit arithmetic. 

We also implemented a version of the rANS Coder which can be loaded as a Python module \footnote{https://github.com/benlcb/rANS-python-module}.

\section{Profiling Results} \label{benchmarks}
\subsection{Compression Rates}

Table \ref{fig:compressionrates} compares the compression performance of DLIC to various established compression algorithms. The dataset used is the 2019 Compression.cc mobile dataset\footnote{http://clic.compression.cc/2019/challenge}. The dataset contains a random selection of mobile phone camera pictures. Vloss referenced in Table \ref{fig:compressionrates} describes the final categorical crossentropy validation loss of the respective network and is indicative for the compression performance of this network, excluding header information. In addition, varying the network size can also be used to trade compression speed against compression rate. Our largest model beats JPEG 2000 by 19\% and FLIF by 7\% respectively. 

\begin{table}[h]
    \centering
\begin{tabular}{ |r|c| } 
 \hline
 \multicolumn{2}{|c|}{Our Model}\\
 \hline
 \# Parameters & vloss in bpp  \\ 
 \hline
 $\sim  100,000$ & 3.5995 \\
$\sim  350,000$ & 3.6211 \\
 $\sim 1,350,000$ & 3.5360  \\
  $\sim 5,330,000$ & 3.5317 \\
 \hline
\end{tabular}
\begin{tabular}{ |r|c| } 
 \hline
 Codec & bpp  \\ 
 \hline
 PNG &  4.2741\\
 HEVC &  4.2701\\
  JPEG 2000 &  4.1986\\
  FLIF &  3.7850\\
  \textbf{Ours} & \textbf{3.5317}\\
 \hline
\end{tabular}
\vspace{0.25cm}
    \caption{Compression rate in bpp for various codecs}
    \label{fig:compressionrates}
\end{table}

\subsection{Runtimes}

\begin{figure}[htbp]
\centerline{\includegraphics[width=\linewidth]{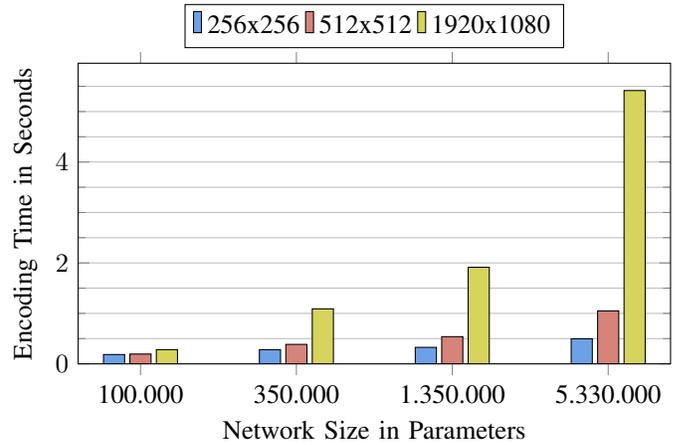}}
\caption{Compression speeds in seconds of different frame sizes by different neural networks sizes, for both encoding and decoding.}
\label{fig:runtimes}
\end{figure}

 Figure \ref{fig:runtimes} shows encoding speeds for different neural network sizes as well as image sizes. The results shown are averaged times over 10 runs. Decoding times do not vary more than 4\% from encoding times. The results were measured on a workstation with a Ryzen 5 2600 CPU and a GTX 1070 Ti GPU running Ubuntu 20.04. While achieving a run time of less than one second, the smallest network with 100.000 parameters could still outperform JPEG 2000 by 17\% and FLIF by 5\%, as shown in \ref{fig:compressionrates}.

\begin{figure}[htbp]
\centerline{\includegraphics[width=\linewidth*0.92]{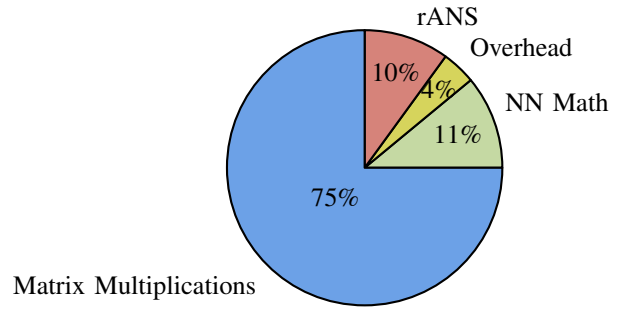}}
\caption{Time spent by task. NN Math consists of bias, softmax and activation function computations. The neural network used for this analysis has about 350,000 parameters. Note that an increase of the number of parameters in the neural network will also increase the relative time spend on matrix multiplications.}
\label{piechart}
\end{figure}

In Figure \ref{piechart}, we provide a runtime breakdown of time spent on the graphics card by type of task. The breakdown can be used to estimate performance on more modern hardware and potential for further optimization: Since 75\% of the compute time is spend performing matrix multiplications, this time will be reduced directly proportionally to the performance increase of newer GPUs.

\section{Domain Specific Application}

A central feature of DLIC is the neural network which computes the PDF used with the entropy coder (see Figure \ref{decode_step}). The network can be trained on domain specific data to learn the features characteristic to the domain data. This process could, in theory, lead to more accurate PDFs and in turn to higher compression rates. We investigate this claim by adopting DLIC to MRI scans.

\subsection{Data}
We collected 537 GB of MRI scans from one practitioner. Because this data included overview images, crops and annotated images, we only used scans which fulfill the following criteria for this paper:
\begin{itemize}
  \item Resolution no smaller than 256 x 256 Pixel
  \item At least 25 slices in one scan
  \item Grayscale scans only
  \item Tags (metadata fields) space between pixels (28,30), space between slices (18,88) and slice thickness (18,50) are present
  \item The bit depth is 12 bit 
\end{itemize}

After filtering the data by the above criteria 109.9 GB over 7771 scans  remained. A random 80/20 split was applied resulting in a training set with 6237 scans (88.2 GB) and a validation set with 1534 scans (21.7 GB).  

\subsection{Considerations}
\label{secB:considerations}

There are several aspects to be taken into account when compressing MRI scans as opposed to gray scale photographs. 
\subsubsection{3D Dependencies}
When performing an MRI scan, a volume of - usually - organic tissue is recorded. In practice, this volume is saved in the form of consecutive planar image files representing slices (e.g. encoded as JPEG) or video files (e.g. AVC). This means that we can conceive both a 2D as well as a 3D DLIC-coder to compress MRI scans. We expect the 3D coder to achieve better compression rates, since it can take 3D dependencies into account. 

\subsubsection{Higher Bit Debth}

The pixels in the volume represent the density of hydrogen atoms in the scanned tissue. This information is captured in one 12-bit channel, leading to 4096 possible shades of gray. We refer to these shades of gray as colors. A wider range of possible colors means that we now have to compute more probability values. This either means to increase the size of the neural network output to 4096 or to implement an alternative solution such as a mixture of logistics.  

\subsubsection{Metadata}
Since the volume represents an actual physical object, metadata is saved alongside the MRI scan to accurately describe the relationship between pixel data and real world. 

\subsubsection{Color Frequencies}
\begin{figure}[htbp]
\centerline{\includegraphics[width=70mm]{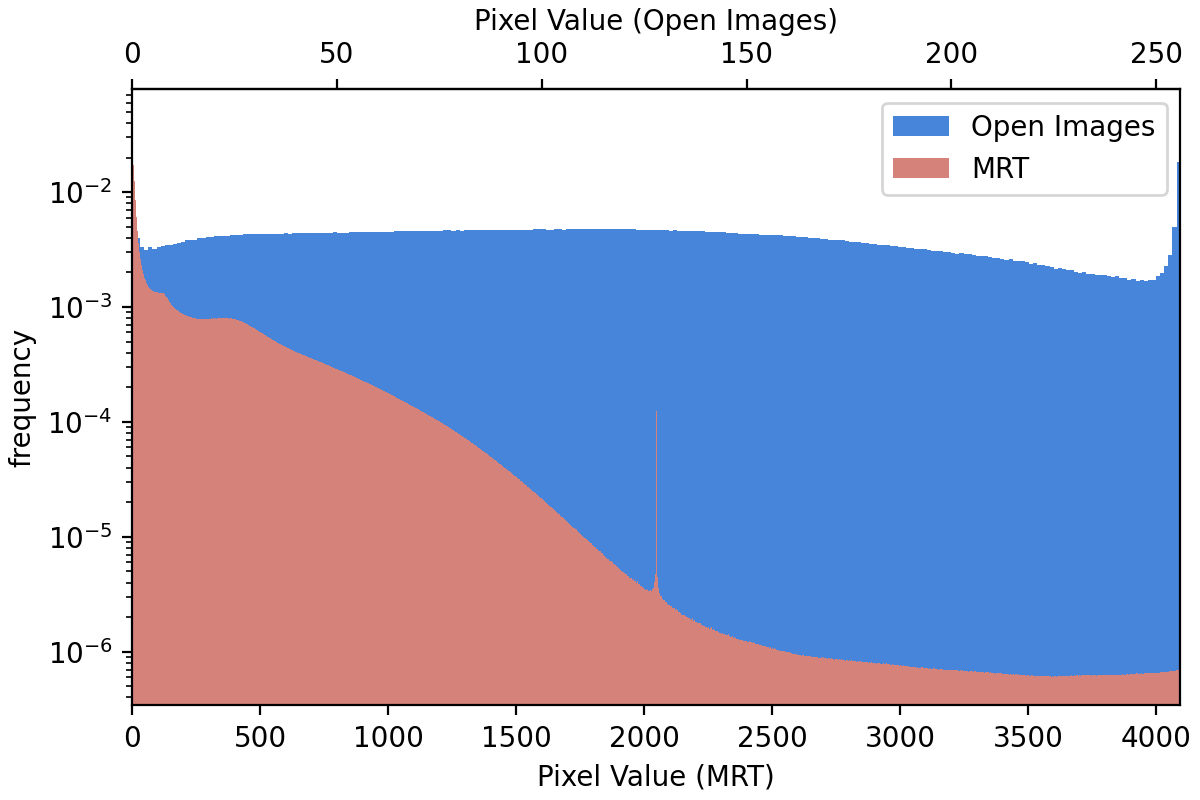}}
\caption{Pixel Value Frequencies by Dataset. Note the logarithmic scale for frequency values. The difference in frequency of occurrence of colors is much larger for MRI images (red) compared to Open Images images (blue). }
\label{histogram}
\end{figure}

Figure \ref{histogram} show the color frequencies in a set of photographs (Open Images Validation Set) and our MRI dataset. The frequencies of the different colors are highly disproportionate in MRI images, varying by a factor of $1*10^4$. In contrast, frequency of colors in the Open Images dataset only vary by a factor of 2. 

\subsection{Implementation}

To account for these various differences we adopted the coder design. 

\subsubsection{3D Dependencies}
To include 3D dependencies into the density estimation, we designed a 3D window. Our statistical and experimental analysis revealed that the dependencies between slices are not as strong as initially expected. The directly adjacent pixels to the target pixel are stronger correlated than the pixel directly above or below and the correlation is diminishing quickly with every additional layer. Consequently we choose a 2-layered window. Figure \ref{3d_window_and_wavefront} (right) shows a 3D window with 3 layers. 

\subsubsection{Higher Bit Depth}
To accommodate the higher bit depth, we also considered a discretized mixture of logistics. The approach did not converge during training, and our final model has 4096 output layer neurons. 

\subsubsection{Metadata}
Our experiments showed that including metadata as a feature can improve the compression rate of DLIC by up to 1.2\% in some cases. The metadata that we used for our compression includes space between pixels (28,30), space between slices (18,88) and slice thickness (18,50), where the numbers refer to the corresponding DICOM\footnote{DICOM is a container format which is typically used to store MRI scans.} tag. We store the metadata uncompressed.

\subsubsection{Color Frequencies}
Adopting the training process to color frequencies might seem unnecessary, since a good density estimation on frequent pixels is desired, even if it comes with a marginal drop in compression rates for very infrequent pixels. The difference in frequency is so large for our DICOM image set, that the performance on colors with low frequency was not acceptable. To address this issue, we upsampled rare pixel values into the standard training set. In addition, we weighted the loss based on the frequency of a pixel values. 

\subsubsection{3D Wavefront}

\begin{figure}[htbp]
\centerline{\includegraphics[width=21mm]{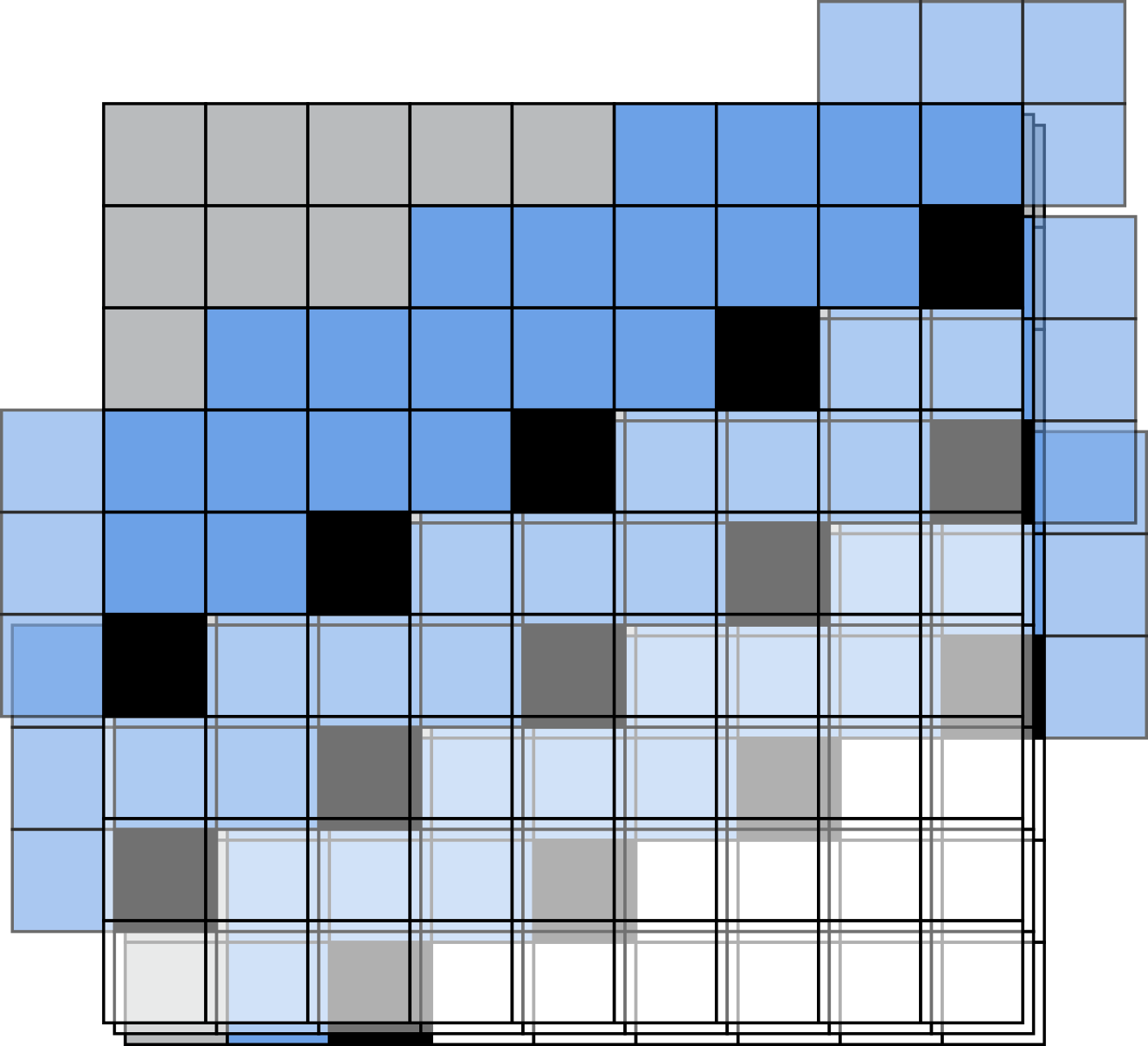}\hspace{10mm}\includegraphics[width=29mm]{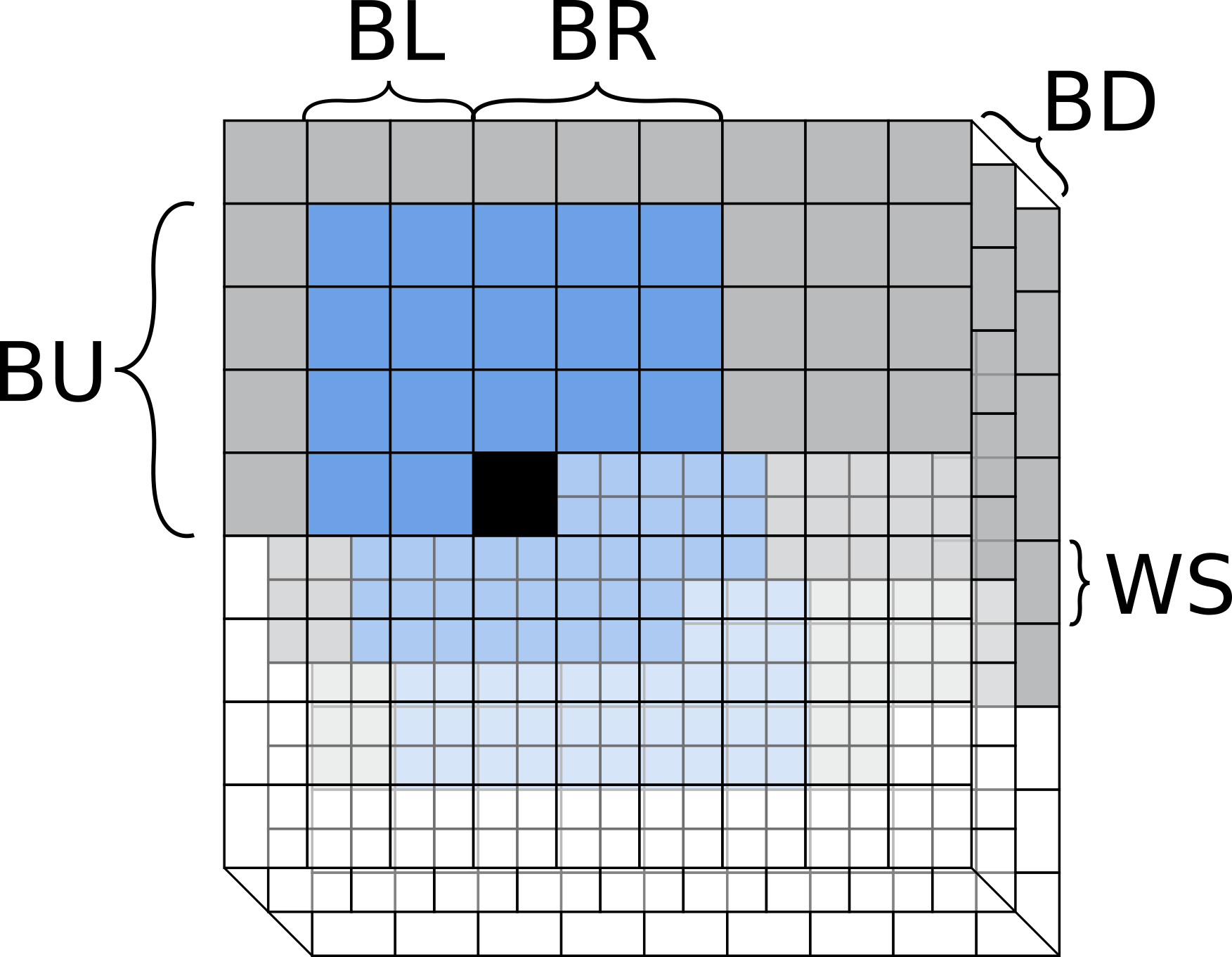}}
\caption{\textbf{Left:} 3D Wavefront. Visualized is a wavefront over 3 layers for a 3D window size of 3-by-3-by-2 pixel. Black pixels are pixels to be decoded, blue pixels are used in a neighborhood window, gray pixels are already decoded. Multiple 2D Wavefronts are overlapped in such a fashion that the pixels from the layer below can be used in the 3D window of the layer above. All black pixels shown in the illustration are decoded at the same time. In practice, the wavefront is not limited to 3 layers, but can be extended to any amount of layers supported by the hardware constraints.
\textbf{Right:} 3D Window design parameters. The top layer includes the pixel of interest, marked in black. The window in the layers below are shifted. This is possible since the wavefront approach can be extended into 3D space.}
\label{3d_window_and_wavefront}
\end{figure}

To further parallelize the coding process for 3D data, we extended the 2D wavefront approach by one dimension. Our 3D wavefront is visualized in Figure \ref{3d_window_and_wavefront} (left). We overlap 2D wavefronts to decode all layers at once. The wavefronts are shifted in each layer in such a way that the pixels used in the neighborhood window in the layer above are already decoded. If the second layer of the 3D window is not shifted down ($WS = 0$, see Figure \ref{3d_window_and_wavefront}), then the wavefront is not shifted and all layers can be decoded at the same time. If the wavefront needs to be shifted, we can calculate the expected speedup by determining the number of wavefronts added for each additional frame. 

\subsection{Results and Evaluation}

\subsubsection{Compression Rates}
A comparison of compression rates can be seen in Table \ref{fig:comparison}. DLIC adopted to MRI images appears superior to JPEG 2000, PNG and FLIF in terms of bpp. Due to the nature of MRI data, a network size of 92 Million parameters was needed to achieve these results. Numerous experiments were done to reduce the network size, such as reducing the resolution for values that occur with a low probability, but description of these is out of the scope of this paper. 

\begin{table}[h]
    \centering
\begin{tabular}{ |l|c|c| } 
 \hline
 Method & bpp train set & bpp valid. set \\ 
 \hline
 TIFF (LZW) & 9.7483 & 9.5664 \\
 PNG (compression = 3) & 8.4937 & 8.3320 \\
 HEVC (12 bit, 3D) & 6.6140 &  6.6038 \\
 HEVC (12 bit, 2D) & 6.4901 &  6.4796 \\
 JPEG (2000, Lossless) & 5.6996 & 5.5864\\
 \textbf{Ours (2D)}$^\dagger$ & \textbf{5.5369} & \textbf{5.5347} \\
 FLIF\cite{flifpaper} & 5.6096 & 5.5016 \\\hline
 \textbf{Ours (3D)}$^\dagger$ & \textbf{5.4534} & \textbf{5.4476} \\
 \hline
 \multicolumn{3}{l}{\footnotesize{${\dagger}$Loss of 92 million parameter network.}}
\end{tabular}
\vspace{0.25cm}
    \caption{Compression rates of different methods on MRI image data}
    \label{fig:comparison}
\vspace*{-0.5cm}
\end{table}

\subsubsection{Coding Duration}

Table \ref{fig:mri_coding_times} shows estimated decoding times for a typical MRI scan. The 3D wavefront approach parallelizes the decoding of the different frames, making the last frame available shortly after the first frame. 

\begin{table}[h]
    \centering
\begin{tabular}{ |r|c|c|c|c| } 
 \hline
 \# Parameters & Scan Size & First Frame (s) & Coding Complete (s) \\ 
 \hline
$\sim  350,000$ & 256x256x35 &  0.2799 & $0.3374$  \\
 $\sim 1,350,000$ & 256x256x35 & 0.3273 & $0.3946$ \\
  $\sim 5,330,000$ & 256x256x35 &  0.4975 & $0.5997$ \\
    $\sim 17,500,000$ & 256x256x35 & 1.1821  & $1.4250$ \\
     $\sim 92,000,000$ & 256x256x35 & 5.9286  & $7.1203$ \\
    \hline

\end{tabular}
\vspace{0.25cm}
    \caption{Estimated Coding Times MRI Scans}
    \label{fig:mri_coding_times}
\end{table}

\subsubsection{Subjective Performance}

\begin{figure}[htbp]
\vspace*{-0.6cm}
\centerline{\includegraphics[width=85mm]{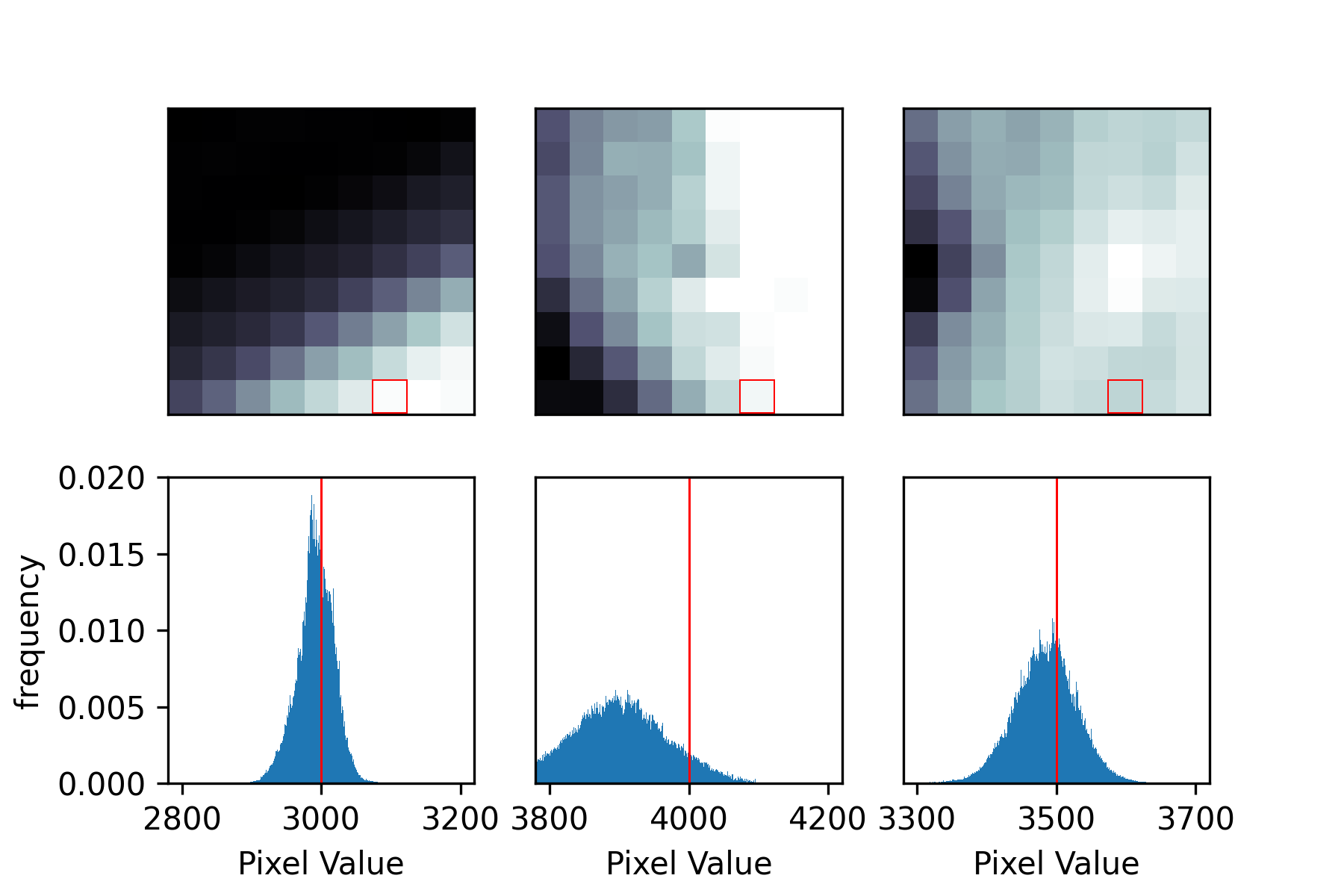}}
\centerline{\includegraphics[width=85mm]{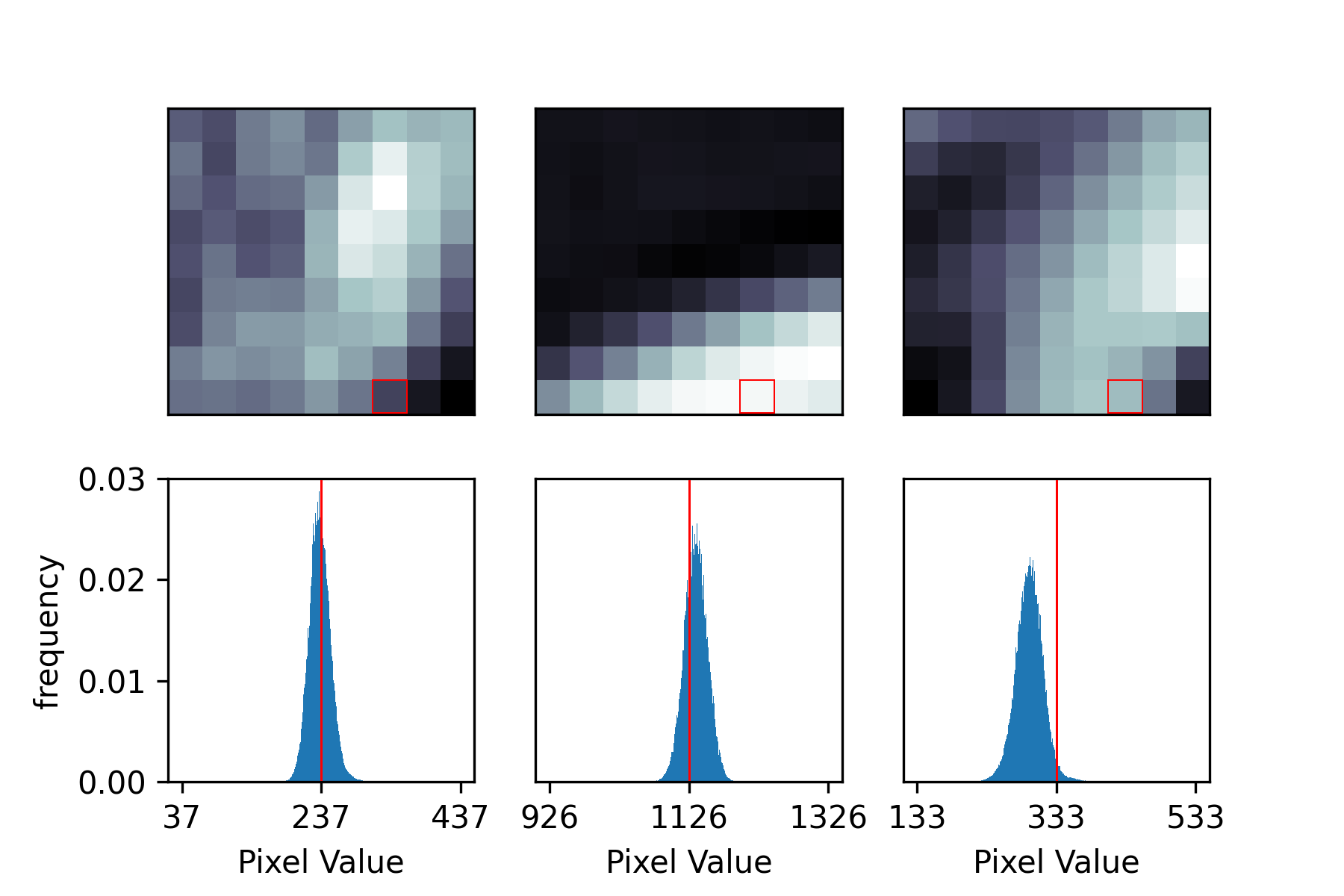}}
\caption{Visualisation of predicted PDFs, \textbf{high values (top)} and \textbf{low values (bottom)}. The top row of each figure shows a visualization of the pixel neighborhood windows. The colors in the window are not to scale due to the large color range. A 9-by-9 window was used; the target pixel is located in the last row, third to last column (marked by a red border). The bottom row of each figure shows the estimated PDF and the target pixel value. The PDFs shown are always centered around their respective target pixel value.}
\label{fig:sub_vis_combined}
\end{figure}

Figure \ref{fig:sub_vis_combined} show a direct visualization of the predicted PDF of the neural network.
The neural network learns a logistic-like distribution around the target value. As discussed in Section \ref{secB:considerations}, the quality of prediction on darker colors is better due to the larger number of samples.

\section{Summary and Conclusion}

We presented a strategy to accelerate lossless, deep neural network based image encoding. Results show that small networks could be sufficient to compete with state-of-the-art compression algorithms for 2D images. Practical runtimes are achieved by parallelizing the encoding and decoding process. 

By adopting our coder to domain-specific data, we showcase the challenges of adopting neural network based solutions to specific use cases. The properties of MRI scans may significantly increase network size and training requirements to be comparable with the state-of-the-art. 

\bibliographystyle{ieeetr}
\bibliography{citations/biblo} 

\end{document}